\documentclass[12pt]{article}
\usepackage{graphicx}
\usepackage{amssymb,amsmath,amsfonts,palatino,amsthm}
\usepackage{amssymb}
\usepackage{epstopdf}
\usepackage{slashed} 
\newcommand{\f}{\begin{equation}}
\newcommand{\ff}{\end{equation}}

\begin{document}

%%%%%%%%%%%%%%%%%%%%%%%%%%%%%%%%%%%%%%%%%%%%%%%%
\title{Lessons from Einstein's 1915 discovery of general relativity \\}
\author{Lee Smolin\thanks{lsmolin@perimeterinstitute.ca} 
\\
\\
Perimeter Institute for Theoretical Physics,\\
31 Caroline Street North, Waterloo, Ontario N2J 2Y5, Canada}
\date{\today}
\maketitle
%\vfill

\begin{abstract}

There is a myth that Einstein's discovery of  general relativity was due to  his following beautiful mathematics to discover new insights about nature.  I argue that this is an incorrect reading of the history and that what Einstein did was to follow physical insights which arose from asking that the story we tell of how nature works be coherent.

This is an expanded version of a text that was originally prepared for, and its Polish translation appeared in, the September 2015 issue of "Niezbednik inteligenta".  An extract also appears in the Fall/Winter 2015/2016 edition of {\it Inside the Perimeter.}
\end{abstract}

\tableofcontents

\newpage

%\section{Introduction}

\section{The lessons of general relativity}

The focus on Einstein's discovery of general relativity brings to mind two questions:

\begin{itemize}

\item{}What can we learn from Einstein's 1915 discovery of general relativity about how science works?

\item{}Are there lessons to be drawn from Einstein's successes and failures that can help our search for a deeper unification?

\end{itemize}

According to popular accounts of the scientific method, such as Thomas Kuhn's {\bf The Structure of Scientific 
Revolutions}\cite{Kuhn}, theories are invented to describe phenomena which experimentalists have previously discovered.  When a theory accounts for all the experimental results, it is considered a success.  But, sometimes a new experimental discovery is made which was not predicted by the theory, making it anomalous.  Then, a new theory is invented to explain the anomaly.   It must, of course, also account for all the old experiments.  When the new theory gives us a correct description of all the experiments, we take it as a replacement for the old theory.  

This simple schema does not apply to general relativity.  All the characteristic phenomena that general relativity describes were unknown in 1915 when Einstein published his theory.  These include the expanding universe, black holes, light bending in gravitational fields, gravitational lenses, time slowing down in gravitational fields, gravitational waves, dark energy.  Not only were these phenomena not yet observed in 1915, most of them had not even been thought about. The fact that a century later, all of these are well confirmed is a triumph unmatched by any other theory in the history of science.

The only competitor might be quantum mechanics but, even though that theory is equally triumphant, it has to be admitted that a large number of characteristic quantum phenomena were known to experimentalists before quantum mechanics was formulated in 1925.

Some people point to the shift in Mercury's perihelion as a case of an anomaly that general relativity explained.  The problem with this is that virtually nobody except Einstein thought this phenomena needed a new theory to explain it.  The bulk of astronomical opinion was that this shift could be accounted for either by a new planet or by more precise calculations of the way the planets' gravitational fields perturb each others orbits.

What should have been clear to anyone who followed physics was that Newton's gravitational theory required revision in the light of special relativity.  But why not introduce a field theory for gravity within the framework of special relativity?  This is the route several of Einstein's contemporaries took and, until Eddington's 1919 observation of light bending by the sun,  this straightforward theory was consistent with all the data.  Why did Einstein ignore this obvious option in favour of a truly radical step?

The idea that a new theory can be invented to challenge a standing theory, even in the absence of experimental anomalies, was suggested by Paul Feyerabend in his book {\bf Against Method}\cite{PKF}.  Feyereband suggests that this strategy can succeed when two things happen: the new theory suggests novel interpretations of existing experiments and these novel interpretations in turn suggest new experiments that distinguish it from the old theory, which would not otherwise have been thought of or seemed significant.    This was certainly the case with general relativity.  The fact that was reinterpreted was the equality of gravitational and inertial mass, which was an accident from the point of view of Newtonian theory and its special relativistic extension, but which becomes a necessary consequence of Einstein's new equivalence principle.   This holds that you cannot tell the difference between an effect of gravity and an effect of inertia.   This succeeded because the equivalence principle had immediately testable consequences such as light bending and the perihelion shift that distinguished it from a special relativistic theory of gravity.  

But how exactly did Einstein perform the seemingly miraculous feat of inventing a theory that correctly describes phenomena that had not yet  even been observed?  There is a myth which is usually trotted out to answer this query, which is that Einstein was a lone genius who followed beautiful mathematics to discover his great theory.  Genius, inspired by aesthetics.  Mathematics as a tool of prophecy.  

No one was more responsible for spreading this myth than Einstein himself, who described in several essays and popular talks in the 1920's and later how he followed a trial of mathematical beauty to his discovery of general relativity.   As Einstein wrote in his autobiographical notes,

`{\it `I have learned something else from the theory of gravitation: no collection of empirical facts, no matter how comprehensive, can ever lead to the formulation of such complicated equations . . . [they] can only be found by the discovery of a logically simple mathematical condition that completely, or almost completely, determines the equations.  Once one has those sufficiently strong formal conditions one requires only little knowledge of facts to set up a theory\cite{auto}.}

In the last twenty years historians have been doing a careful job of studying what Einstein actually did during the eight years of hard, often frustrating work it took him to create this theory\cite{historians}.  Their verdict is that, as often happens, the myth Einstein invented, which has been so avidly accepted and spread by many physicists, is just that.  By careful analysis of his notebooks and letters, page by page, equation by equation,  the historians have put together the true story of Einstein's road to general relativity.  And it was very different from the myth.  

I was alerted to this conclusion by Professor Jurgen Renn, who is head of the Max Planck Institute for the history of science in Berlin, and one of the most important of the scholars who study Einstein's work.  I owe the view point I present here to what I learned from him and other historians such as John Satchel and Julian Barbour. 

There is no doubt that Albert Einstein was a singular creator who contributed more  than anyone else to the invention of both quantum theory and relativity theory.  In the history of physics he is comparable only to Isaac Newton.  When one reads the papers and books of the other great physicists, such as Galileo and Kepler, Maxwell, Bohr, Heisenberg, Schroedinger, Dirac, I daresay many physicists will get the feeling of  understanding who they are.  They are extremely good scientists, but not different in kind from the very best of our contemporaries.  But Newton and Einstein are different, after many years of study, during which time I have often reread them,  I still find their unerring, surefooted ability to penetrate right to the heart of things to uncover the secrets of nature incomprehensible.  

But even so, the myth is not helpful.  To begin with, Einstein didn't work alone.  He was, for sure, a leader who followed his own compass, to his own ideas and insights.  But he had collaborators, competitors and friends with whom he was in continual communication as he worked.

Why is it important to tell the correct story of what Einstein accomplished? If we hope to emulate his achievements it will help to understand how he got to his great theories.  There are also lessons be learned from the failures of such a great mind.

Einstein was neither very well educated in mathematics, nor very good at it.  He depended on friends such as Marcel Grossman to explain to him the mathematics on which general relativity is based.  And he depended on other friends, such as Michael Besso, to find the correct interpretation of the mathematics.  Indeed, contemporaries noted that there were many colleagues who were much better at mathematics, such as John von Neumann. 

Unlike Newton,  Einstein did not invent any of the mathematics he used to express his new theories.   General relativity employs mathematics that was advanced for the time-the mathematics of curved surfaces and general geometries which had been developed by mathematicians in the second half of the 19th Century.  Einstein was the first physicist to use this new approach to geometry to describe physical systems.  But he followed the tuition of Marcel Grossman in learning and applying the mathematics.

Indeed, Einstein was not very good at using this new mathematics.  Once he had written down and published the equations of general relativity, solutions which describe simple examples were quickly found.  These describe very symmetric situations such as spherically symmetric stars and homogeneous, expanding universes.  To derive these solutions are now homework exercises in undergraduate courses in general relativity.  But Einstein didn't find any of these simple solutions, indeed there is no evidence he even looked for them.  They were found by others within weeks of his papers being published.  

What Einstein excelled at was physical intuition and insight.  His path to general relativity was brightly illuminated by a simple physical idea: the equivalence principle.   This simple idea is revolutionary because it precludes a gravitational field theory consistent with special relativity.  

I had a very happy day about fifteen years ago when I visited Jurgen Renn in Berlin and he showed me images of the notebooks in which Einstein had created general relativity.  What impressed me was that Einstein was using the same techniques all physicists use to grasp the essential features of a phenomena they want to model.  These are the development of approximate expressions, together with the playful creation of simple examples and models.  These are the tools every physicist is taught, which they employ throughout their career, first, to do their homework and, later, to make progress in their research.  

The mathematics Einstein used may appear beautiful to some who study it, but what is going on in Einstein's notebooks was not beautiful.  It was hardheaded and pragmatic.  When you dine at a  fancy restaurant you may be impressed by the aesthetic presentation of a dish as it is brought to the table.  But this is only the last step, just as the freshness of the ingredients as they come from the farm is only the first step.  In between, hidden in the kitchen, it is all just hard, practical work.  Mistakes are made, but these, ideally, never leave the kitchen.  In Einstein's kitchen-his notebooks-it was no different.

Why did Einstein weave a myth around his creation of general relativity?  What was his motive for telling a fable about the role of mathematical beauty in his creation of general relativity?

The reason may be that he was making propaganda to promote interest in work he was doing to follow up on general relativity.  This was aimed to go beyond general relativity to a theory he hoped would be his masterpiece, a unified theory of all phenomena, incorporating not just gravity but also electromagnetism.   He called this the unified field theory.  His ambition for it was huge for, not only was it to describe all the forces in nature, it was to replace quantum mechanics.  For Einstein was a disappointed parent when it came to the second of his theoretical children.  He had early in the 20th Century done more than anyone to give birth to quantum theory.  But when that theory was finally put into final form in the 1920's-by others-he was very unhappy with the outcome.   Quantum theory made uncertainty and probability fundamental and Einstein rejected this.  He sought a deeper description which would give a complete and deterministic description of each and every individual phenomena.  He hoped this would be his unified field theory.  

The problem was that Einstein had no physical insights to guide his search for this next unification.  He had no new physical principles to propose, no new thought experiments to provoke his thinking.  Unlike each of his prior successes, special relativity, photons, Brownian motion and general relativity, Einstein was working without guidance from his formidable physical intuition.  He was running, as Jackson Brown sings, on empty.  

In the absence of ideas and insights about nature, Einstein fell back on mathematics as his guide.  He constructed a myth about how mathematical beauty had been prophetic for his invention of general relativity and he attempted to use it to justify his forays into unified field theory.   

Einstein's search for a unified field theory failed, and the roots of this failure are his embrace of mathematical beauty as a guiding principle.  Over the thirty-five years between 1920 and his death in 1955 Einstein attempted many versions of a unified field theory.  He tried higher, hidden dimensions, they failed.  He tried more general versions of curved geometries beyond the geometry used in general relativity.  They all failed to produce a useful unification.  

Einstein already understood by 1922 that the hypothesis that there are extra, hidden dimensions could not give a unification of the forces.   The reason is that if the extra dimensions have dynamical geometries-as they must if they are to be unified with how general relativity describes the four dimensions we know-this implies the properties of the elementary particles will be unstable. For example, the values of the electric charges of the elementary particles will vary in time.  We know to impressive accuracy that this doesn't happen.  

The stability of the electric charges and other properties of the elementary particles means that, if they are understood in terms of geometry, those geometries cannot evolve in time.  But to the extent that this is true, these geometries are not unified with gravity as general relativity describes it.

As Einstein wrote to his friend Paul Ehrenfest, {\it ``It is anomalous to replace the four dimensional continuum by a five dimensional one and then to subsequently tie up artificially one of those five dimensions in order to account for the fact that it does not manifest itself."}\cite{1922}. 

At the time Einstein also failed to convince many to follow him in the search for the unified field theory.  His previous work, on quantum theory and special and general relativity had provoked quick interest from the leading physicists and mathematicians of the day.   This time was different.   Apart from a handful of assistants, only a  few, such as Erwin Schroedinger, followed him into the swamp of unified field theories.  

Most of the younger people rejected his leadership and followed instead quantum physics, condensed matter physics (which Einstein had also pioneered) quantum field theory and particle physics.  This led to triumph on triumph, ending with the standard model of particle physics.  Written down in 1973, this theory did, partly and incompletely, unify all the forces save gravity.  And it explains all the results of experiments not involving gravity.  The standard model is, like quantum  theory and general relativity, based on a few simple physical ideas, such as the role of gauge symmetry, chiral symmetries and broken symmetries.

\section{Following Einstein's path}

The story I've just sketched suggests that Einstein left us two very different and largely incompatible legacies.  From the early Einstein, we learn to work from physical principles and thought experiments, and to develop our  physical intuition.  After 1919, Einstein began to ignore his unique strengths and proclaimed and followed a different methodology, based on mathematical 
aesthetics\footnote{We should not expect to find a single date dividing the young from the older Einstein.  Like any complex person, Einstein was pulled in different directions.  We begin to see the older Einstein's reliance on mathematics over physics as early as the late teens, while there are traces of the early Einstein at least up to the $EPR$ paper of 1935.}.    

In the search of physics beyond the standard model, many of us have taken the  second, failed path Einstein pioneered, of relying on mathematical beauty for inspiration when new insights into physics failed to appear.   And, you know what? We have failed, just as the later Einstein did.  Indeed, some of the same ideas that Einstein failed with have been revived and have failed again for us, just as definitively.  

There have been a few new ideas, such as strings and their higher dimensional avatars-called branes.  Thousands of theorists have spent decades studying these ideas, and there is not yet a single connection with experiment.  The reason is that these objects have apparently to live in  extra, hidden, dimensions.  There are beautiful theories which describe these strings and branes moving around in the higher, hidden dimensions.  And they fail to make contact with nature for exactly the same reason Einstein's explorations of higher dimensional geometries fail. They cannot explain why the higher dimensional geometries are frozen as is needed to keep the elementary particles stable.  (In the technical literature this is called the problem of stabilizing moduli.)

In the 1960's Roger Penrose proved theorems which tell us that the universe cannot be static or eternal, but must have evolved from singularities.   As Penrose has pointed out, these same theorems apply to the higher dimensional unifications Einstein studied and they apply equally to the higher dimensional geometries string theory is based on.  They imply that these unifications cannot be stable.  

The idea that all particles and forces come from oscillations of a string is a reasonable physical hypothesis.  It might be right, it might be wrong.  The problems begin when the mathematics tell us that such strings can only move consistently in a nine dimensional space. This blocks the easy, direct application to physical phenomena in our three dimensional world.   What are we to do with the six extra dimensions?  How are they to be arranged, what geometry will they choose and what will keep them stable?  There are an infinite number of choices for the extra dimensions, and no principle guides their choice. As a result, string theory makes no definite predictions that might be tested by experiment.

\section{Going beyond the standard model: which legacy to follow?}

Einstein taught us to seek novel physical principles.   But this doesn't always work out.  Since the 1970's the search for a new unification beyond the standard model has been to a large extent guided by two principles.

\begin{enumerate}

\item{\bf Naturality.}

\item{}{ \bf Unification through symmetry.}

\end{enumerate}

The first idea is that {\it the dimensionless parameters of a truly fundamental theory should be few and these should not require fine tuning to match experiments,} i.e they should be ``natural" numbers like combinations of low powers of small integers and $\pi$ rather than terribly big or small numbers.

The standard model of particle physics has about 30 adjustable dimensionless parameters.  And many of them are very tiny or very big.   Examples are the cosmological constant, the masses of the neutrinos, the masses of the Higgs particle, the masses of the heavier quarks.  Thus the standard model does not satisfy the principle of naturality.  The hope is that the standard model can be explained by a deeper theory that is natural.

The search  for this deeper theory is guided by the second idea, which is that {\it the smaller we probe into nature,
the more unification we should find. }

There have been circumstances where naturality and increased symmetry were powerful heuristics.  The former played a role in our understanding of chiral symmetry breaking in the physics of pions.  Maxwell was led to the his equations by asking for greater symmetry in the equations.  But as clues for how to go beyond the standard model they have so far proved unreliable.  Naturality has so far not led to explanations of either the Higgs scale or the cosmological constant.  Hypotheses such as supersymmetry and grand unification, as compelling as they are, have not so far been confirmed.

At this juncture we might wonder why the search for principles has not worked out for us as well as it worked for Einstein.

Notice that these are different kinds of principles than those that Einstein invented, such as the principle of equivalence and the principle of the relativity of inertial frames.  Those principles are directly about nature.  They constrain, and can be falsified by, individual experiments. They require no mathematics to express them: their contents can be entirely captured in a verbal description of an experiment.  Historians talk of ``thought experiments",  but in fact the principles invented by the young Einstein referred to genuinely doable experiments.

The two principles just stated are somewhat different.
You can not express or directly test them in a single experiment.  They are more meta-principles.  They are formal principles that constrain theories.   

Both of these are, in essence, aesthetic ideas.  And they express a particular kind of aesthetics: mathematical aesthetics.  They are not beautiful ideas about nature, such as the equivalence principle or the principle of the relativity of inertial frames.   The beauty of those older ideas is not lost when we express them in words.  In contrast, naturality and unification through symmetry are ideas which are best expressed by referring to the equations that specify a theory.  They express the hope that the mathematics in which we express our theories will be beautiful.

The research guided by these two principles has  been relying on aesthetics in place of more directly physical principles.   

If we could, we would begin with new physical principles such as the principles of relativity and equivalence,  which imply particular symmetries. But no such principles have been proposed.  So we are left only with the idea that we are looking for a  unification would be expressed mathematically by embedding the symmetry group of the standard model in some unknown bigger symmetry group or algebra. 

There are two problems with the strategy of reaching for some bigger symmetry, without a physical principle to tell us what that larger symmetry should be.  It is not terribly proscriptive.  There are several ways to extend a symmetry, for example, by being embedded in a larger symmetry group, or by quantum deformation, affine extension, supersymmetry.  Then, because the symmetry is not directly expressed in nature it must be broken-explicitly or spontaneously.  Either option introduces additional  free, undetermined parameters.  Thus the strategy of seeking bigger unifications leads to more complexity, ambiguity and more undetermined parameters.

We do have a few examples of theories which explain the same facts the standard model explains, but do so more naturally,  employing fewer very tiny numbers.  This can be done using grand unification, supersymmetry, technicolor, or large extra dimensions.  Each of these theories has its attractive points, and each predicts new particles and new forces.  These have been looked for and so far, unfortunately, not seen. This doesn't mean the theories are wrong, as the new particles may be found at still higher energies.  But it does mean they will not solve the puzzle of unnaturality, because the parameters of these theories themselves need fine tuning to explain why the characteristic phenomena they predict have not been seen.

An example is supersymmetry.  Supersymmetry is at first a  physical principle: all particles come in pairs of bosons and fermions, whose interactions are constrained by a physical transformation which exchanges them.  At first this seems an interesting hypothesis.  Indeed, were there such a supersymmetry between pairs of known particles, that would be very restrictive, and it would give us a lot of insight into how nature is structured.   But there isn't-no such hypothesis has been proposed.  So we are reduced to proposing that each known particle is matched to a so far unknown super-partner.  But no super-partners are observed so the supersymmetry must be broken.   So the super-partners are hidden either by having large masses or by other mechanisms.  

The problem is that if we are allowed to just make up super-partners and then tune parameters to make them invisible,  we get few  constraints on unification-as the standard model itself may be supersymmetrized.  And so can many variants or alternatives to it.  Nor does supersymmetry decrease the number of free parameters-because of the need to break supersymmetry, the number of free undetermined parameters is greatly increased.   As a result we can set the masses of the super-partners as high as we need to be consistent with the failure of the most recent experiments to detect them.  

Originally there was the idea that supersymmetry would have naturally explained the hierarchy problem by constraining the radiative corrections to the Higgs mass.  But many versions of this idea imply super-partner masses that have been excluded by the experiments.  

So the combination of supersymmetry and naturality appears to be ruled out.  Similar fates have befallen other apparently beautiful ideas about unification such as $SU(5)$ grand unification.  That is, to the extent that they are natural they are testable and indeed, in several cases such as supersymmetry, technicolor and grand unification the simple natural versions have been ruled out.  

Similarly, string theory is an idea about unification that vastly  increases the number of possibilities for particle physics.  So, contrary to what we expected, it seems that mathematically based schemes for unification, imposed in the absence of physical principles, vastly decreases the explanatory power of our theories\footnote{For a counter-example to this, see \cite{Cohl}. which describes a mathematical explanation for the structure of the standard model.}.

The standard model remains the best description we have of the fundamental particles and forces. And it is highly unnatural, i.e. very ugly.  But it has survived when more beautiful theories have failed.  

This may suggest that the parameters of the standard model are not fixed, timeless expressions of deep laws of nature.  Perhaps they are dynamical quantities that evolve on cosmological scales\cite{CNS,LOTC}.  If so, we should ask that the principle that governs that evolution of the parameters should be natural.  But the parameters themselves need not be.

The lesson which we have had to painfully learn-a second time-is that mathematical beauty is not prophetic of discoveries about nature.  There is no royal road to understand nature that does not centrally employ insights and hypotheses about nature.  Mathematics can be very helpful when it is used to develop ideas about nature but, in the absence of insights about nature,  it cannot be trusted to show the way forward.

%Another example of a possible new principle that has been very influential in the search for a deeper unification, beyond the standard model, is the {\it principle of naturality}.   The laws of physics have some numbers in them, like the masses of the quarks and electron and other particles or the strengths of the forces.  Here is an idea about mathematical beauty which is easy to state: the ratios of these numbers should not be too big or too small. For if they were, it would require explanation.  Tiny numbers and huge numbers are ugly, indeed they can be said to be unnatural.  The principle that they not appear in the statement of the fundamental laws is called naturality, and it is clearly an aesthetic principle.  

\section{The search for new principles}

The contrast between the successes of the young Einstein and the failures of the later Einstein teach us a lesson.  This lesson is reinforced by the failures of our own efforts the last 40 years to follow the later Einstein in seeking guidance from formal, mathematical principles, rather than physical principles.  The lesson is that the task of formulating a physical principle must come first-only when we have one in hand do we have a basis to look for new mathematics to express the new principle.  

Are there principles that can guide our ongoing search for a more complete understanding of the laws of nature?   

One principle that seems reliable is {\it background independence\cite{BI}.}  This says that the laws of nature should be statable in a form that does not rely on the specification of a fixed geometry of spacetime.  Einstein's theory of general relativity satisfies this principle, and it has been a useful heuristic for the search for quantum gravity.  Background independence can be understood as expressing Leibniz's principle of sufficient reason, which states that there should be a reason for every choice made in the formulation of the laws of nature\cite{PSR}. This underlies the idea that space and time are aspects of relationships among dynamical degrees of freedom.

One implication of this principle is that there can be no fundamental symmetries in the laws of nature.  Every event in the history of the universe must be describable uniquely in terms of the relational degrees of freedom.  This means that the closer we are to a fundamental theory, the {\it fewer} symmetries we should have.  This may be why our search for larger and larger symmetries is no longer  working.  

The principle of background independence implies that the structures that are fixed in the laws of nature can be understood to be the result of dynamical evolution in time.  The paradigmatic example is the geometry of spacetime which is fixed and absolute in Newtonian physics and special relativity, but which general relativity reveals to be dynamical.  

One of the lessons of general relativity is then that it can be a good strategy to seek to identify fixed, apparently arbitrary,  structures in the statements of the laws of nature and replace them by dynamical degrees of freedom which evolve in time by deeper laws.  

Thus, the principle of background independence can be extended to say that the laws of physics should not rely on any fixed structures-whether that is Hilbert space metric or the geometry of the internal space where the degrees of freedom of the elementary particles live.  All such structures should then be the result of dynamics.  

This suggests that the search for laws of nature may end up not with a single choice but with a dynamical principle operating on a landscape of theories\cite{CNS,LOTC}.  Such a dynamical principle might provide a dynamical mechanism for tuning the constants of the standard model that would explain why they have been fine tuned.  That is, making the structure of our physical theory dynamical might give a natural explanation for unnaturality.

The search for quantum gravity has produced one candidate for a new physical principle, which is {\it the holographic principle.}  It was put forward by Gerard 't Hooft\cite{thooft},  based on a desire-one he shares with Einstein-to go beyond quantum mechanics. This says that a model {\it world with gravity can be described as if it were a world without gravity, with one fewer dimension, where that surface theory has one degree of freedom per Planck area.}  This is inspired by thinking about the implications of Bekenstein's discovery that the entropy of physical systems is bounded by the area of a surface that encloses them, in Planck units.   
 
 Leonard Susskind\cite{Lenny} and Juan Maldacena\cite{Malda} have applied the holographic principle to string theory, where it turned out to be extremely illuminating.  It has other applications beyond string theory which suggest it is a truly general principle.  Unfortunately, these so far do not apply to our world, because they require the dark energy be negative when, in nature, it is positive.  Still, this is one of the very best idea we have so far and it shows we can get further if we start with insights and principles, as 't Hooft did. 

My sense is that the hAs so far stated, the holographic principle fails to have the direct physical content of the principles of relativity and  equivalence.  It cannot be expressed or tested in a single experiment.  

We also so far lack  a formulation of the holographic principle which is consistent with the principle of background independence that grounds general relativity\cite{bi-holo}.  The boundary conditions that Maldacena's form of the principle impose-that the spacetime be asymptotically $AdS$, imposes a fixed background and so breaks background independence.  The negative cosmological constant and asymptotically $AdS$ condition describe a subsystem placed in a box.  We need instead a formulation of the holographic principle that applied to a closed universe with a positive cosmological constant, rather than a sub system of the universe in a box defined by a negative cosmological constant.   

%It may be that the fact that we haven't found its deepest expression, consistent with background independence, is behind our failure to find the  right mathematics to express the holographic principle.  

\section{Einstein's unique approach to physics}

This brings us to a last question:   what made the young Einstein different from his contemporaries, which allowed him to make discoveries others couldn't?   With a hesitation due to my appreciation of the subtlety of his thought, here is a tentative answer:

The start of the answer is that Einstein asked different questions than his contemporaries.  Why?  Because he had a deep need to tell a coherent story about the world.  His contemporaries were content to live with  knowledge that is incomplete and to a greater or lessor degree, contradictory or incoherent.

There is nothing wrong with this. Most scientists have other fish to fry-other goals than to seek the greatest coherence in our knowledge of the universe.  Most are content to pick low hanging fruit and advance knowledge incrementally.  Einstein, more than anyone else, did science to satisfy a deep need to understand himself placed in a  coherent universe.  

Because of this he was very alert to situations in which two phenomena which are indistinguishable experimentally have very different explanations.  To him such cases indicate we could understand something more deeply.   His two great principles: that of relativity and that of equivalence are of this kind.

Some people would describe such cases as symmetries.  But the essence has nothing to do with mathematics, even if we might eventually employ mathematics to describe them.  This is not beautiful mathematics, these are cases primarily of beautiful insights into how nature works.  The point is not how beautiful the equations are, it is how minimal the assumptions needed and how elegant the explanations.

Remember, in 1905 Einstein proposed the special theory of relativity by invoking two physical principles-the relativity of inertial frames and the constancy of the speed of light and insisting on their mutual coherence.  He did not invent the fourth dimension or spacetime.  Those came later-proposed by a mathematician.  

Einstein was suspicious of apparent coincidence.  Newton had two notions of mass-inertial mass, or resistance to force, and gravitational mass, or weight.  But the two masses always turn out to be equal.  For everyone else, this equality of gravitational and inertial mass was just an extra condition to be imposed on the equations.  For Einstein, this was a tremendous opportunity to discover a hidden coherence.  Maybe from the right point of view, gravity and inertia are the same. Einstein found that point of view, and that is the key to general relativity.  

This need for coherence drew him to a certain philosophy-that expounded by Ernest Mach and, before him, Leibniz.  They argued that space and time are not absolute, that is fixed and structured without regard to what exists or how it moves.  Instead, to these sages, space and time are relational, so that their properties reflect the positions and motions of the matter in the universe.  This is more coherent, because it explains more with fewer assumptions.  

But Einstein was no ideologue.  He took inspiration from the writings of some philosophers, but he was happy to creatively misunderstand them in the service of the physical insights he was seeking.  Einstein's use of philosophy, like his use of mathematics, is opportunist and pragmatic.  Above all, he seeks coherence in our understanding of nature.   

Einstein succeeded when he was able to formulate a principle or hypothesis about nature, which he, or sometimes others, later expressed in mathematical terms.  He failed when he attempted to use mathematics as a substitute for insight into nature.  You can indeed use mathematics to unify gravity and electromagnetism, in fact I know of at least four ways to do this. But in the absence of a physical insight or principle as to what the unification means, experimentally, the mathematical unification is empty.  

So as we celebrate the birthday of  general relativity let us admire the Einstein who achieved that great step: a pragmatic but determined seeker after coherence, a physicist who had an unmatched  power of insightfully getting to the hidden story at the heart of natural phenomenon.

\section*{ACKNOWLEDGEMENTS}

It is a pleasure to thank  Dennis Overbye,  Abraham Pais, Jurgen Renn, Julian Barbour, John Satchel and Johnny Wheeler, for discussions about Einstein.  I am grateful to Jerzy Kowalski-Glikman and Natasha Waxman for their encouraging suggestions regarding this essay.  

This research was supported in part by Perimeter Institute for Theoretical Physics. Research at Perimeter Institute is supported by the Government of Canada through Industry Canada and by the Province of Ontario through the Ministry of Research and Innovation. This research was also partly supported by grants from NSERC, FQXi and the John Templeton Foundation.


\begin{thebibliography}{99}

\bibitem{Kuhn}Thomas Kuhn, {\bf The structure of scientific  revolutions},
University of Chicago press, 1962.

\bibitem{PKF}Paul Feyerabend, {\bf Against Method}  (New Left Books, 1975).

\bibitem{auto}Albert Einstein, {\it Autobiographical notes}, in Einstein,  Paul Arthur Schilpp. {\bf Albert Einstein Philospher-Scientist} Vol. 7. Open Court Publishing Company, 1949.

\bibitem{historians}Renn, Jurgen, ed. The Genesis of General Relativity: Sources and Interpretations. Vol. 250. Springer Science \& Business Media, 2007;  
Janssen, Michel, and Jurgen Renn. "Untying the knot: How Einstein found his way back to field equations discarded in the Zurich notebook." The genesis of general relativity. Springer Netherlands, 2007. 839-925.
.
\bibitem{1922} Quoted in Abraham Pais, {\bf Subtle is the lord} (New York: Oxford Univ. Press, 1982) p. 334

\bibitem{BI}Lee Smolin The relational idea in physics and cosmology, contribution to a book, Structural realism and quantum gravity, edited by Steve French. Preprint version: The case for background independence, hep-th/0507235.

\bibitem{PSR}Gottfried Wilhelm Leibniz, THE MONADOLOGY, 1698, translated by Robert Latta, availabe at http://oregonstate.edu/instruct/phl302/texts/leibniz/monadology.html; Leibniz, Gottfried Wilhelm (Oxford Philosophical Texts), ed. by Francks, Richard and Woolhouse, R. S., Oxford Univ. Press, 1999; H. G. Alexander, ed., The Leibniz- Clarke Correspondence, Manchester University Press, 1956, for an annotated selection, see http://www.bun.kyoto-u.ac.jp?suchii/leibniz-clarke.html.

\bibitem{CNS}L. Smolin, 1992a.   "Did the Universe Evolve?" 
Class. Quantum Grav.  9,  173-191;   A perspective on the landscape problem, Invited contribution for a special issue of Foundations of Physics titled: Forty Years Of String Theory: Reflecting On the Foundations, DOI: 10.1007/s10701-012-9652-x  arXiv:1202.3373

\bibitem{LOTC}L. Smolin {\it The Life of the Cosmos}, 
1997 from Oxford University Press (in the USA),
Weidenfeld and Nicolson (in the United Kingdom) and Einaudi Editorici
(in Italy.); 
Time Reborn, Aprill 2013, Houghton Mifflin Harcourt, Random House Canada and Penguin (UK);  Roberto Mangabeira Unger and Lee SmolinThe Singular Universe and the Reality of Time, Cambridge University Press, November 2014.  

\bibitem{Cohl}Cohl Furey, {\it Generations: Three Prints, in Colour}, arXiv:1405.4601, JHEP 10 (2014) 046; 
{\it Unified Theory of Ideals,}
 arXiv:1002.1497, Phys. Rev. D 86, 025024 (2012).


\bibitem{thooft}G. 't Hooft, gr-qc/9310006, hep-th/0003004.

\bibitem{Lenny}L. Susskind, J. Math. Phys. 36 (1995) 6377, 

\bibitem{Malda}J. Maldacena,
Adv. Theor. Math. Phys. 2 (1998) 231.

\bibitem{bi-holo}Fotini Markopoulou and Lee Smolin,  {\it  Holography in a quantum spacetime, }hep-th/9910146, Oct, 1999;  Lee Smolin, {\it The strong and weak holographic principles,} hep-th/0003056, Nucl.Phys. B601 (2001) 209-247.

\end{thebibliography}
\end{document}